\documentclass[]{iopart}
\usepackage{iopams}
\usepackage[]{graphicx}
\usepackage{color}

\newcommand{\IF}{\vartheta}
\newcommand{\erfc}{\textrm{erfc}}
\newcommand{\ha}{\hat{a}}
\newcommand{\hb}{\hat{y}}
\newcommand{\low}{\textrm{\tiny low}}

\begin{document}

\title[]{Bayesian estimate of the zero-density frequency of a Cs fountain}
\author{D Calonico, F Levi, L Lorini and G Mana}
\address{INRIM - Istituto Nazionale di Ricerca Metrologica, Str.\ delle Cacce 91, 10135 Torino, Italy}
\ead{g.mana@inrim.it}

\begin{abstract}
Caesium fountain frequency-standards realize the second in the International System of Units with a relative uncertainty approaching 10$^{-16}$. Among the main contributions to the accuracy budget, cold collisions play an important role because of the atomic density shift of the reference atomic transition. This paper describes an application of the Bayesian analysis of the clock frequency to estimate the density shift and describes how the Bayes theorem allows the a priori knowledge of the sign of the collisional coefficient to be rigourously embedded into the analysis. As an application, data from the INRIM caesium fountain are used and the Bayesian and orthodox analyses are compared. The Bayes theorem allows the orthodox uncertainty to be reduced by 28\% and demonstrates to be an important tool in primary frequency-metrology.
\end{abstract}

\submitto{Metrologia}
\pacs{02.50.Cw, 02.50.Tt, 06.20.Dk, 07.05.Kf, 06.30.Ft}


\section{Introduction}
Atomic fountains depend on laser cooling of caesium atoms down to a temperature of 1 $\mu$K or an even lower value; this, together with the use of a single microwave cavity to implement the Ramsey separated-field spectroscopy, allowed the accuracy of primary frequency standards to be improved by more than one order of magnitude. Today, fountains realize the second with a relative accuracy ranging from $3\times 10^{-16}$ to $10\times 10^{-16}$ \cite{M06,M06_bis,NISTF1,syrtef1,ptbf1,nplf1,nict,nmij} and, in the last ten years, allowed the uncertainty of the Atomic International Timescale (TAI) unit to be reduced from $1\times10^{-14}$ to $5\times 10^{-16}$.

In Cs fountain clocks, cold collisions occur between the ultra-cold atoms with consequent perturbation of the atomic energy levels and shift of the atomic reference frequency. This shift, is proportional to the atom density and it must be carefully evaluated to correct the clock frequency. Since it is a major component of the uncertainty budget, it has been the subject of many theoretical and experimental studies \cite{Williams,Winands}. It depends on the collisional dynamics during the atom ballistic flight, which is related to the way the atom cloud is prepared before launch. When an atom cloud, captured into a magneto-optical-trap, is launched, its high initial density limits the collision energy and the density shift is strongly dependent on the temperature and the mixture ratio \cite{Winands}. However, if, as in fountains operation, the atom cloud is prepared with direct molasses capture or molasses expansion after magneto-optical-trap capture \cite{M06}, the collision energy is high during the whole flight and the shift has a negligible sensitivity to the atom temperature. Additionally, fountains are operated in a way that the collision energy is between 7 $\mu$K and 10 $\mu$K range and that the atomic sample is almost an evenly quantum mixture of the hyperfine eigenstates $|F=3, m_F=0 \rangle$ and $|F=4, m_F=0 \rangle$. In this case, the density shift has a negligible dependence on the mixture ratio and it is linearly dependent on density through a negative coefficient \cite{Williams}.

In a previous paper \cite{CLLM08}, we used the Bayesian inference \cite{sivia,jaynes} to estimate the collisional coefficient, consistently with the constraint of a negative value. This estimate can be used to extrapolate the fountain frequency to the zero-density value. In the present paper we illustrate how to perform the extrapolation, given two or more frequency measurements at different atomic densities, irrespectively of the value of the collisional coefficient value, but consistently with its negative sign. From a mathematical viewpoint, the problem is to find the best-fit line through two or more points, with the constraint that the regression coefficient is negative.

Our goal is twofold. First, we want to illustrate a non-trivial application of the Bayes theorem and the relevant data analysis. Second, we want to assess the viability a density shift estimation consistent with the constrain of a negative collisional coefficient and to test its performances. A Bayesian approach is important because the shift has the same magnitude as the measurement noise. Therefore, despite the fact that the shift is a linear function of the atom density having a negative regression coefficient, measurement results having the low-density frequency lower than the high-density one are relatively common. A linear extrapolation to zero density from these data is clearly meaningless. On the contrary, a Bayesian inference makes it possible to deal with this situation, thus avoiding physical absurdity and, consequently, reducing the extrapolation uncertainty.

After the problem statement, in section \ref{statement}, the Bayesian solution is given, first for two frequency measurements and, then, for list of frequency measurements. Eventually, in section \ref{sbs-rem}, the Bayesian analysis is applied to data collected in a test measurement. All the symbolic and numerical calculations have been performed with the aid of Mathematica \cite{math}.

\section{Experimental techniques}
The density shift is commonly evaluated by means of a differential measurement approach. The fountain is operated alternating low with high atom density and the frequency of a hydrogen-maser is measured using the fountain in the two configurations. By virtue of its frequency stability, in both short and medium time scales, the maser is used as a flywheel oscillator: the comparison of the two measured frequencies cancels the maser frequency and allows density shift to be evaluated. As reported in \cite{Williams}, if the ratio between the atomic density and the total number of detected atoms is assumed constant, we can state that the density shift is proportional to the number of detected atoms. The differential measurement provides a collisional coefficient which can be used to extrapolate the frequency data to zero density.

The actual experimental practice in the various laboratories differs in the way the atom clouds of different densities are prepared and in the durations of the fountain operation at low and high density. The extrapolations techniques of the fountains of the Physikalish Technische Bundesanstalt (PTB-CsF1), of the National Physical Laboratory (NPL-CsF1), and of the National Institute of Standards and Technology (NIST-F1) differ only by the driving parameter and the time scheduling of the differential measurements. At the Syt\`emes de R\'ef\'erence Temps Espace, a rapid adiabatic passage technique is used, which ensures very accurate evaluation of the collisional coefficient because the atom density is precisely set at one value and at its half.

At INRIM, we vary the atom density through the loading time of the magneto-optical trap. To evaluate the density shift, the fountain is operated alternately at high and low atomic density; the 70 ms and 300 ms loading times provide respectively the low and high density configurations. The ratio between the number of detected atoms in the two configurations ranges from three to four. Since in the low density configuration stability is poor and resolution is limited by the atom shot noise, the fountain is operated alternating about 21000 s in the low density configuration and about 6000 s in the high density one. The hydrogen maser frequency is then extrapolated to zero density by
\begin{equation}\label{extra}
 {\hat y} = y_1 - \frac{y_2-y_1}{x_2-x_1} \,x_1,
\end{equation}
where $\hat{y}$ is the sought zero density frequency, $y_1$ and $y_2$ are the frequency in low and high density conditions, and $x_1$ and $x_2$ are the number of atoms detected at the low and high density, respectively. The extrapolation is carried out for each  $(y_1,y_2)$ pairs. The total duration of each run is 27000 s; this ensure a good rejection of systematic effects -- fluctuations of the hydrogen maser frequency, magneto-optical trap, and atom detection efficiency -- which could bias the frequency extrapolation.

By neglecting the $\sigma_x$ uncertainty of the atom-density measurements -- an acceptable omission provided $\sigma_x \ll \sigma_y/|y_2-y_1|$ -- the uncertainty of (\ref{extra}) is
\begin{equation}\label{s0}
 \hat{\sigma} = \frac{\sqrt{x_1^2+x_2^2}}{x_2-x_1} \; \sigma_y ,
\end{equation}
where $\sigma_y$ is the uncertainty of the frequency measurements. Deviations from a linear relationship and an imperfect rejection of long term effects are estimated to contribute to the uncertainty by 20\% of the  density-shift value. Here, this non-statistical contributions to the uncertainty have not been considered, though they take part of the total uncertainty budget of atomic clocks. Besides, being the statistical contribution of the same order of magnitude as the shift, model errors do not contribute significantly.

\section{Zero density extrapolation}

\subsection{Statement of the problem}\label{statement}
In the simplest case, we want to perform a linear regression analysis of two clock-frequency, at low and high atom densities, given the prior knowledge that the slope of the sought line is negative. In particular, we are interested in the value of the intercept of the regression line. The data are assumed normally distributed about $ax_i+b$ with the same standard deviations $\sigma_y$, that is,
\begin{equation}\label{Py}
 P_\nu(y_i|a,b) = \frac{1}{\sqrt{2\pi}\sigma_y} \exp \big[ -\frac{(y_i-ax_i-b)^2}{2\sigma_y^2} \big] .
\end{equation}
The joint probability density of the data is
\begin{equation}\label{gauss}\fl
 P_\nu(y_1,y_2|a,b) =
 \frac{1}{2 \pi\sigma_y^2} \exp \big[ -\frac{(y_1-ax_1-b)^2 + (y_2-ax_2-b)^2}{2\sigma_y^2} \big] .
\end{equation}
The question is how to account for the $a<0$ constraint and to infer a consistent value for $b$.

Here and in the following, we will use the notation $P_r(r_i|s_j)$ to indicate the probability density that the quantity $r$ has the particular value $r=r_i$ if the parameter $s$ in the probability distribution has the particular value $s=s_j$. Irrelevant conditionals, such as $x_i$ and $\sigma_y$ in (\ref{Py}), will be dropped.

\subsection{Bayesian solution}
The Bayesian approach takes account of $a<0$ into the probability density $P_{ab}(a_0,b_0)$ of the regression coefficient and intercept values before the measurement result is know; in the absence of any additional information let us assume that
\begin{equation}\label{prior}
 P_{ab}(a_0,b_0) = \IF(-a_0) ,
\end{equation}
where $\IF(.)$ is the Heaviside function. According to the Bayes theorem the post-data probability density of the $a$ and $b$ values is
\begin{eqnarray}\label{post}\fl
 P_{ab}(a_0,b_0|y_1,y_2) &\propto &P_\nu(y_1,y_2|a_0,b_0)P_{ab}(a_0,b_0) \nonumber \\
 &\propto &\exp \big[ -\displaystyle\frac{(y_1-a_0x_1 - b_0)^2 +(y_2-a_0x_2-b_0)^2}{2\sigma_y^2} \big]\IF(-a_0) ,
\end{eqnarray}
where a meaningless normalization coefficient has been omitted. This probability density embeds both the $a<0$ constraint and the information delivered by the data.

\begin{figure}[b]\centering
\includegraphics[width=7cm]{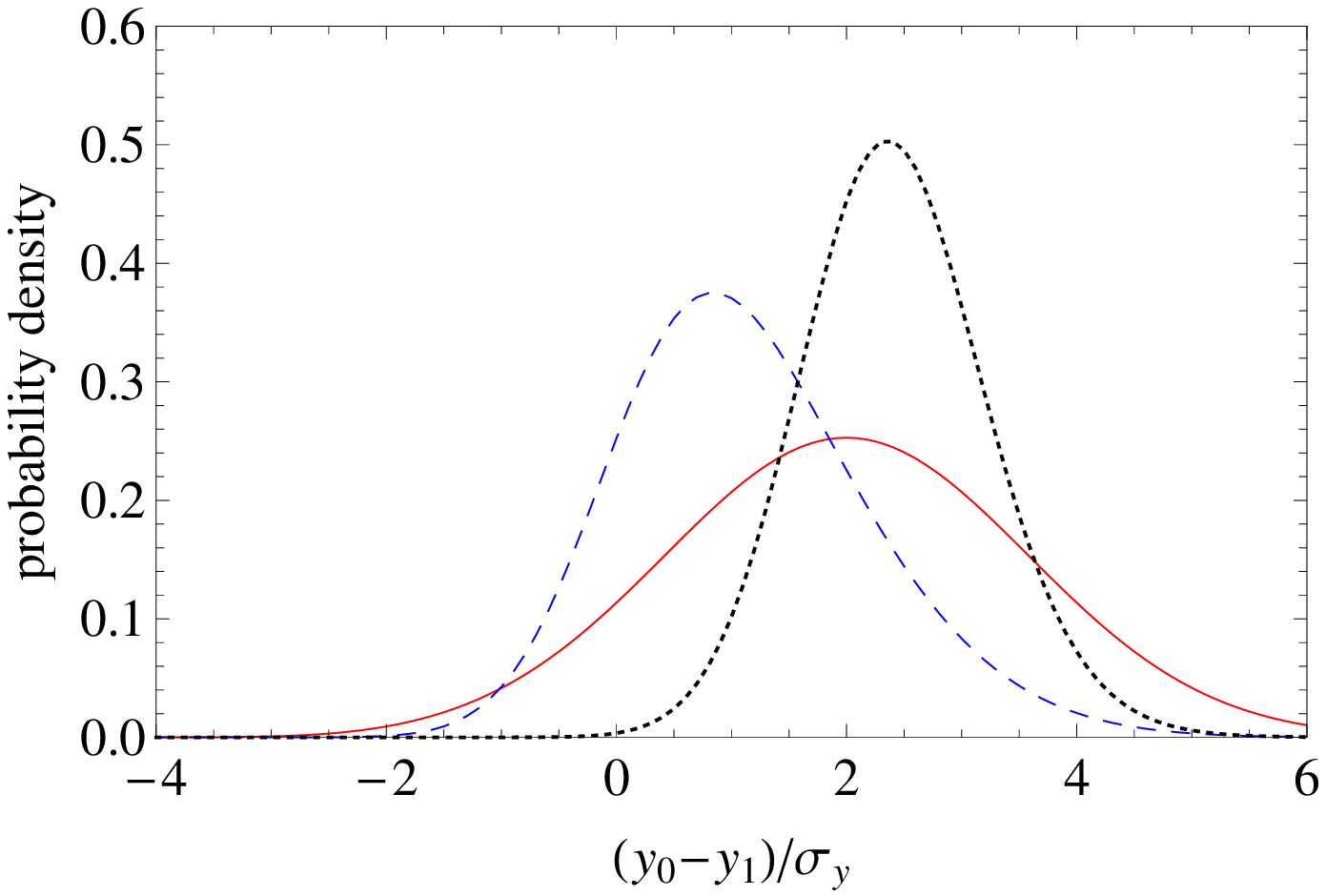}
\caption{Probability density of the intercept values; $x_2/x_1=3$, solid (red) is $(y_2-y_1)/\sigma_y=-4$, dashed (blue) is $(y_2-y_1)/\sigma_y=0$, dotted (black) is $(y_2-y_1)/\sigma_y=+4$.} \label{pdf}
\end{figure}

To obtain the probability density of the zero-density frequency, irrespectively of the regression coefficient, $a_0$ is integrated out from (\ref{post}) by marginalization. Hence \cite{math},
\begin{eqnarray}\label{marginal}
 P_b(b_0|y_1,y_2) &= &\int_{-\infty}^{0} P_{ab}(a_0,b_0|y_1,y_2) \; \rmd a_0 \nonumber \\
 &\propto &\exp \big[- \frac{(b_0-\hb)^2}{2\hat{\sigma}^2} \big]
 \erfc \big[ -\frac{b_0-{\tilde y}}{\sqrt{2}{\tilde \sigma}} \big] ,
\end{eqnarray}
where $\hb$ and $\hat{\sigma}^2$ are the intercept of the line through the data and its variance -- given by (\ref{extra}) and (\ref{s0}),
\begin{equation}
 {\tilde y} = \frac{x_1y_1 + x_2y_2}{x_1+x_2} ,
\end{equation}
is the mean of the measured frequencies weighed by the relevant atomic densities,
\begin{equation}
 {\tilde \sigma}^2 = \frac{x_1^2+x_2^2}{(x_1+x_2)^2} \,\sigma_y^2
\end{equation}
is the uncertainty of $\tilde y$, and $\erfc(.)$ is the complementary error function. The post-data probability density is shown in Fig.\ \ref{pdf}. The Gaussian factor is the probability density we obtain by a classical analysis: it is the probability density of the orthodox extrapolation, having $\hb$ mean and $\hat{\sigma}^2$ variance. The $\erfc$ factor, originating from the application of the Bayes theorem and from marginalization, takes account of the $a<0$ constraint and cuts off the intercept values less than $\tilde y$.

The probability density (\ref{marginal}) is the result of the Bayesian analysis. To convert it into a single numerical estimate, a loss associated with the estimate error must be specified. The optimal estimator minimizes the expected loss over (\ref{marginal}). A loss proportional to the squared or absolute errors indicates the mean or the median, respectively; a constant loss indicates the most probable values. Confidence intervals are easily expressed by integrating (\ref{marginal}) to obtain the cumulative distribution function.

\begin{figure}\centering
\includegraphics[width=7cm]{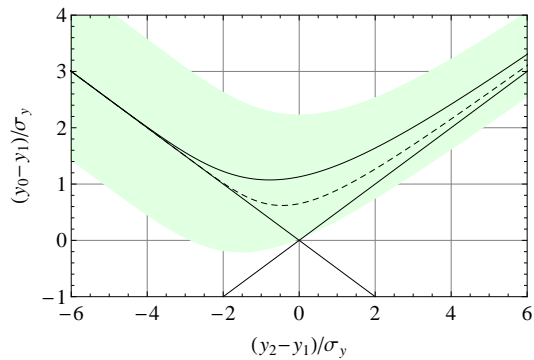}
\caption{Mean and standard deviation of the zero-density frequency, given $\{x_1=1,y_1\}$ and $\{x_2=3,y_2\}$. The left and right sides of the diagram correspond to physical and unphysical values of the collisional coefficient, respectively. The coloured area indicates the extrapolation uncertainty. The straight lines (asymptotic limits) are $b=y_1-(y_2-y_1)/2$ (best-fit line intercept) and $b=(y_1+y_2)/2$ (sample mean). Dashed line is the mean when the same $\{1,y_1\}$ and $\{3,y_2\}$ pairs are observed three times.} \label{intercept}
\end{figure}

Figure \ref{intercept} shows the mean and standard deviation of the zero-density frequency. In the limit case when $(y_2-y_1)/\sigma_y \ll 0$, the mean tends to the $\hb$ intercept of the line through the data, but with larger uncertainty. This is explained by observing that, via marginalization, the Bayesian analysis takes account of all the possible regression coefficients. The understanding of the $(y_2-y_1)/\sigma_y \gg 0$ limit requires some training of our intuition. Let us observe that \cite{math}
\begin{equation}
 \lim_{(y_2-y_1)/\sigma_y \rightarrow\infty} P_b(b_0|y_1,y_2) =
 \frac{1}{\sqrt{\pi}\sigma_y} \exp \bigg[ -\displaystyle\frac{(b_0-\bar{y})^2}{\sigma_y^2} \bigg] ,
\end{equation}
where $\bar{y}=(y_1+y_2)/2$. Therefore, the intercept mean approaches the sample mean. In this case, the data are inconsistent with the a priori knowledge on $a$, but we did not allow room for such inconsistency when formulating the problem and, though at least one of the data is clearly wrong, we do not know which it is. Consequently, the best we can do is to fit the data with a line satisfying the $a<0$ constraint. The sample mean corresponds to $a=0$; the Bayesian inference is slightly greater because it accounts for all the $a<0$ values. When $(y_2-y_1)/\sigma_y \gg 0$, the extrapolation uncertainty is smaller than the uncertainty of the extrapolation when $(y_2-y_1)/\sigma_y \ll 0$. The reason is that, when $(y_2-y_1)/\sigma_y \rightarrow \infty$, we did not question the $\sigma_y/\sqrt{2}$ uncertainty of the sample-mean, regardless of data inconsistency. Finally, we observe that, when $(y_2-y_1)/\sigma_y \approx 0$, the mean is always greater than $y_1$ and smoothly connects the $\hb$ and $\bar{y}$ asymptotes.

When $(y_2-y_1)/\sigma_y \approx 0$, the Bayesian analysis seems to overestimate the zero-density frequency, which, from an orthodox analysis, is expected to be quite near $y_1$. The supposed overestimation is due to the scarce information delivered by the data and the use of (\ref{prior}): with a uniformly negative regression coefficient, a vertical regression is, a priori, as probable as a horizontal one and the Bayes theorem accounts for both possibilities. In the appendix A, we give the result of a Monte Carlo simulation which confirms that (\ref{prior}) describes correctly the pre-data probability density, when no information, apart $a<0$, is available. To summarize the results, we used the mean and the standard deviation as calculated from (\ref{marginal}). However, when $(y_2-y_1)/\sigma_y \approx 0$, the post-data probability density is distorted; in this case, the most probable value is somewhat smaller than our estimate.

The role of the assumed pre-data probability density of the regression coefficient is also shown by considering the Bayesian estimate of the zero-density frequency given a single data pair, for instance, $\{x_1=1,y_1=0\}$. In this case, the marginal probability density of the best-fit line intercept is the improper distribution
\begin{equation}
 P_b(b_0|x_1) \propto \erfc\bigg( -\frac{b_0}{\sqrt{2}\sigma_y} \bigg) ,
\end{equation}
where roughly all the frequencies greater than the measured value are equally probable. However, as an increasing number of data become available, the post-data distribution is dominated by the likelihood function; $P_{ab}(a_0,b_0)$ becomes irrelevant, and we are led to the same conclusion irrespectively of the prior probability density.

\subsection{Extension to repeated observations}
To extend the previous analysis to $N$ measurement pairs $\{x_i,y_i\}$, we must rewrite the joint probability density of the data as
\begin{equation}\label{many}
 P_\nu(\bi{y}|a,b) = \prod_{i=1}^N N(y_i|ax_i+b,\sigma_i^2) ,
\end{equation}
where $\bi{y}$ is the vector of the measured frequencies, $N(y_i|\nu_i,\sigma_i^2)$ is a normal distribution with $\nu_i$ mean and $\sigma_i^2$ variance, and we assumed the data independent and identically distributed. Hence, the joint post-data probability density of the regression coefficient and zero-density frequency is
\begin{eqnarray}\label{post-multi}
 P_{ab}(a_0,b_0|\bi{y}) &\propto &P_\nu(\bi{y}|a,b) \IF(-a_0) \\
 &\propto &\exp \bigg[ -
 \frac{ \left( \begin{array}{cc} a_0-\ha & b_0-\hb \end{array} \right)
        C_{ab}^{-1}
        \left( \begin{array}{c} a_0-\ha \\ b_0-\hb \end{array} \right) }
 {2} \bigg] \IF(-a_0) , \nonumber
\end{eqnarray}
where, as it is shown in the appendix B, $\ha$, $\hb$, and $C_{ab}$ are the least-squares estimates and covariance matrix of $a$ and $b$. Eventually, in the same way as in (\ref{marginal}), the parameter $a_0$ is integrated out of the problem by marginalization. There is no additional insight in integrating (\ref{post-multi}) analytically, but it is must be noted that, when $\ha/\sigma_a \ll 0$, where $\sigma_a$ is the least-squares uncertainty of $\ha$, then $E(b)\rightarrow \hb$, where $E(b)$ is the $b$ mean over the $P_b(b_0|x_i,y_i)$ distribution. On the contrary, when $\ha/\sigma_a \gg 0$, then $E(b)\rightarrow \bar{y}$, where $\bar{y}$ is the sample average of the data. In the general case, the smaller the uncertainty of the least-squares line is, the closer is the extrapolated frequency $E(b)$ to $\hb$, if $\ha<0$, and to $\bar{y}$, if $\ha>0$. This is shown in Fig.\ \ref{intercept} for the particular case when the same $\{1,y_1\}$ and $\{3,y_2\}$ pairs are observed three times.

\subsection{Discussion of results}
We can also process the data pairs sequentially. In this case, we start with (\ref{post}) based on the first data pair. This probability density substitutes for (\ref{prior}) as the pre-data probability density in the analysis of the second data pair. When this procedure is repeated up to the last pair, we obtain the same post-data distribution as that obtained with the one-step approach. To better understand the Bayesian analysis, let us consider three identical data-pairs, where $(y_2-y_1)/\sigma_y$ is close to zero. By starting with (\ref{post}), given the first data pair, and by repeating the analysis given the second and, then, given the third, we note that the Bayesian extrapolation changes each time a new pair is made available. This result is in striking contrast with the orthodox analysis where, if we use three identical data pairs, the extrapolation uncertainty is reduced, but the extrapolated value remains the same.

This apparent paradox is explained by observing that, when applying the Bayes theorem for the first time, the only piece of information is that the regression coefficient value is between zero and minus infinity with equal probability. Hence, the extrapolation is biased towards a frequency value much greater than $y_1$. In the second iteration, additional information is available, namely the result of the first measurement. Both pieces of information are synthesized in the post-data probability density (\ref{post}), which substitutes for the pre-data probability density (\ref{prior}) and limits extrapolation to a neighbour of the classical one. In each subsequent application of the Bayes theorem, we update the pre-data probability density, thus further reducing the discrepancy between orthodox and Bayesian extrapolations. The Bayes theorem, by prescribing that the regression coefficient is negative and that all its negative values -- including those arbitrarily large -- have the same prior probability, infers that the classical extrapolation statistically underestimates the zero-density frequency. However, when several measurement results are available, and all are consistent with $a<0$, Bayesian extrapolation approaches to the classical one.

In the absence of informative data, the sensitivity of the Bayes theorem to prior information, synthesized in the pre-data probability density of the measurand, may be disappointing. This could discourage the use of Bayesian methods, in view of an apparent lack of objectivity. However, a seminal paper by R.\ T.\ Cox \cite{cox} demonstrates that, in order to make consistent inferences, it is necessary to resort the Bayes theorem. If we give it up, because we are adverse to make an estimate depending on the prior information, we put our results at the risk of contradictions.

\subsection{Estimation of the hydrogen maser drift} \label{sbs-rem}
Given its good stability during time intervals from hours to weeks, the hydrogen maser is a common choice as a frequency flywheel for the differential measurements. It is also used as a transfer oscillator when the fountain is used to evaluate the TAI time unit. However, the hydrogen-maser frequency drifts linearly; this drift is usually evaluated and removed by means of orthodox statistical techniques. Since the drift is stable for long time intervals, we can use the knowledge accumulated in the past fountain run to exploit a full Bayesian simultaneous evaluation of the drift and zero-density frequency, with the use of a two-dimensional linear regression.

This requires a slight modification of the sampling distribution of the data, which are therefore assumed normally distributed about $ax_i+b+ct_i$, where $x_i$ and $t_i$ are respectively the density and epoch related to the frequency value $y_i$, $a$ is the collisional coefficient, $c$ is the hydrogen maser drift and $b$ is the frequency value extrapolated to zero density and the epoch $t=0$.  Hence, the joint probability density of the data becomes
\begin{equation}
 P_\nu(\bi{y}|a,b,c) = \prod_{i=1}^N N(y_i|ax_i+b+ct_i,\sigma_i^2) .
\end{equation}
The drift of the hydrogen maser frequency is known; therefore, its pre-data probability density is
\begin{equation}
 P_{c}(c_0) = N(c_0|\tilde c,\sigma_c) ,
\end{equation}
where $\tilde c \pm \sigma_c$ is its pre-data estimate. Eventually, the joint post-data probability density of the model parameters is
\begin{equation}\fl
 P_{abc}(a_0,b_0,c_0|\bi{y}) \propto \prod_{i=1}^N N(y_i|a_0x_i+b_0+c_0t_i,\sigma_i^2) N(c_0|\tilde c,\sigma_c) \IF(-a_0)
\end{equation}
and the post-data probability densities of each individual parameter irrespectively of the others are obtained by marginalization. Then, the marginal post-data probability density of the zero-density frequency is
\begin{equation} \label{postdataeq}
 P_{b}(b_0|\bi{y}) = \int_{-\infty}^{+\infty} P_{abc}(a_0,b_0,c_0|\bi{y})\; \rmd a_0\; \rmd c_0
\end{equation}

\begin{table}[h]\centering  \label{tabdata}
\caption{Clock frequency {\it vs.} atom density. Measurement have been performed alternately at low and high densities at equal 0.313 day intervals; time increases from top to bottom and, then, from left to right. $\rho_{\low}$ and $\nu_{\low}$ are the mean low-density and -frequency, $\sigma_\nu=3.9\times 10^{-15} \nu_{\rm Cs}$ is the mean uncertainty of frequency data.}\vspace{2mm}
{\tiny
\begin{tabular}{llllllll}
\hline
$ $\\
  $\frac{x_1}{\rho_\low}$   &$\frac{y_1-\nu_\low}{\sigma_\nu}$ &$\frac{x_2}{\rho_\low}$ &$\frac{y_2-\nu_\low}{\sigma_\nu}$ &$\frac{x_1}{\rho_\low}$  &$\frac{y_1-\nu_\low}{\sigma_\nu}$ &$\frac{x_2}{\rho_\low}$ &$\frac{y_2-\nu_\low}{\sigma_\nu}$ \\
$ $\\
\hline
$0.960(32)$ & $-4.09(84)$ & $3.42(10)$ & $-1.27(1.15)$  & $0.966(24)$ & $+0.19(82)$ & $3.28(11)$ & $-2.09(1.15)$ \\
$0.980(32)$ & $-1.94(84)$ & $3.43(10)$ & $-3.01(1.15)$  & $0.998(24)$ & $-0.21(82)$ & $3.53(11)$ & $+0.87(1.15)$ \\
$0.970(32)$ & $-1.99(84)$ & $3.42(10)$ & $-1.94(1.15)$  & $1.044(24)$ & $-0.09(82)$ & $3.74(11)$ & $-0.58(1.15)$ \\
$1.001(32)$ & $-2.60(84)$ & $3.53(10)$ & $-3.19(1.15)$  & $1.067(24)$ & $+0.20(82)$ & $3.73(11)$ & $-1.99(1.15)$ \\
$0.979(32)$ & $-1.80(84)$ & $3.40(10)$ & $-0.40(1.15)$  & $1.048(24)$ & $+0.77(82)$ & $3.60(11)$ & $-1.91(1.15)$ \\
$0.974(32)$ & $-1.11(84)$ & $3.47(10)$ & $-0.30(1.15)$  & $1.019(24)$ & $+0.93(82)$ & $3.58(11)$ & $+1.64(1.15)$ \\
$1.027(32)$ & $-1.83(84)$ & $3.57(10)$ & $-3.62(1.15)$  & $1.027(16)$ & $+0.62(82)$ & $3.64(12)$ & $-2.14(1.15)$ \\
$0.963(32)$ & $-2.78(84)$ & $3.37(10)$ & $-1.73(1.15)$  & $1.045(16)$ & $+1.93(82)$ & $3.55(12)$ & $-0.79(1.15)$ \\
$0.942(32)$ & $-3.75(84)$ & $3.27(10)$ & $-2.24(1.15)$  & $1.057(16)$ & $-0.38(82)$ & $3.68(12)$ & $+1.82(1.15)$ \\
$0.960(32)$ & $-1.83(84)$ & $3.38(10)$ & $-1.86(1.15)$  & $1.025(16)$ & $+0.52(82)$ & $3.45(12)$ & $+1.23(1.15)$ \\
$0.942(32)$ & $-2.11(84)$ & $3.30(10)$ & $-2.78(1.15)$  & $1.067(24)$ & $+1.26(1.66)$&$3.63(12)$ & $+2.05(1.15)$ \\
$0.957(32)$ & $-1.87(84)$ & $3.31(10)$ & $-1.76(1.15)$  & $1.043(48)$ & $+3.15(82)$ & $3.42(11)$ & $+2.40(1.12)$ \\
$0.968(32)$ & $-1.38(84)$ & $3.49(10)$ & $-3.80(1.15)$  & $0.922(48)$ & $+1.02(82)$ & $3.25(11)$ & $+2.05(1.12)$ \\
$0.999(32)$ & $+0.52(84)$ & $3.63(10)$ & $+1.97(1.15)$  & $1.017(48)$ & $+1.45(82)$ & $3.62(11)$ & $+1.79(1.12)$ \\
$1.021(32)$ & $+0.50(84)$ & $3.25(10)$ & $-0.48(1.15)$  & $1.039(48)$ & $+1.61(82)$ & $3.52(11)$ & $+1.36(1.12)$ \\
$0.976(32)$ & $-0.89(84)$ & $3.18(10)$ & $-1.71(1.15)$  & $0.949(48)$ & $+1.69(82)$ & $3.21(11)$ & $+1.66(1.12)$ \\
$1.025(32)$ & $-2.02(84)$ & $3.62(10)$ & $-0.63(1.15)$  & $0.988(48)$ & $+3.68(82)$ & $3.72(11)$ & $+1.59(1.12)$ \\
$1.021(32)$ & $-0.52(84)$ & $3.17(10)$ & $+0.34(1.15)$  & $1.048(48)$ & $+2.91(82)$ & $3.53(16)$ & $+2.45(1.12)$ \\
$0.972(32)$ & $-0.35(84)$ & $3.47(6)$  & $-1.40(0.64)$  & $0.970(48)$ & $+4.25(82)$ & $3.12(16)$ & $+4.55(1.12)$ \\
$1.049(32)$ & $-0.32(1.48)$&$3.35(6)$  & $-1.55(0.64)$  & $0.992(48)$ & $+4.31(82)$ & $3.69(16)$ & $+3.12(1.12)$ \\
$0.980(32)$ & $+0.78(1.48)$&$3.18(6)$  & $-0.40(0.64)$  & $1.046(48)$ & $+2.98(82)$ & $2.89(16)$ & $+2.17(1.12)$ \\
$0.956(32)$ & $-1.42(1.48)$&$3.15(14)$ & $+0.05(1.12)$  \\
\hline
\end{tabular}}
\end{table}

\section{Frequency extrapolation in a Cs fountain}
\begin{figure}\centering\hspace{-2mm}
\includegraphics[width=6.6cm]{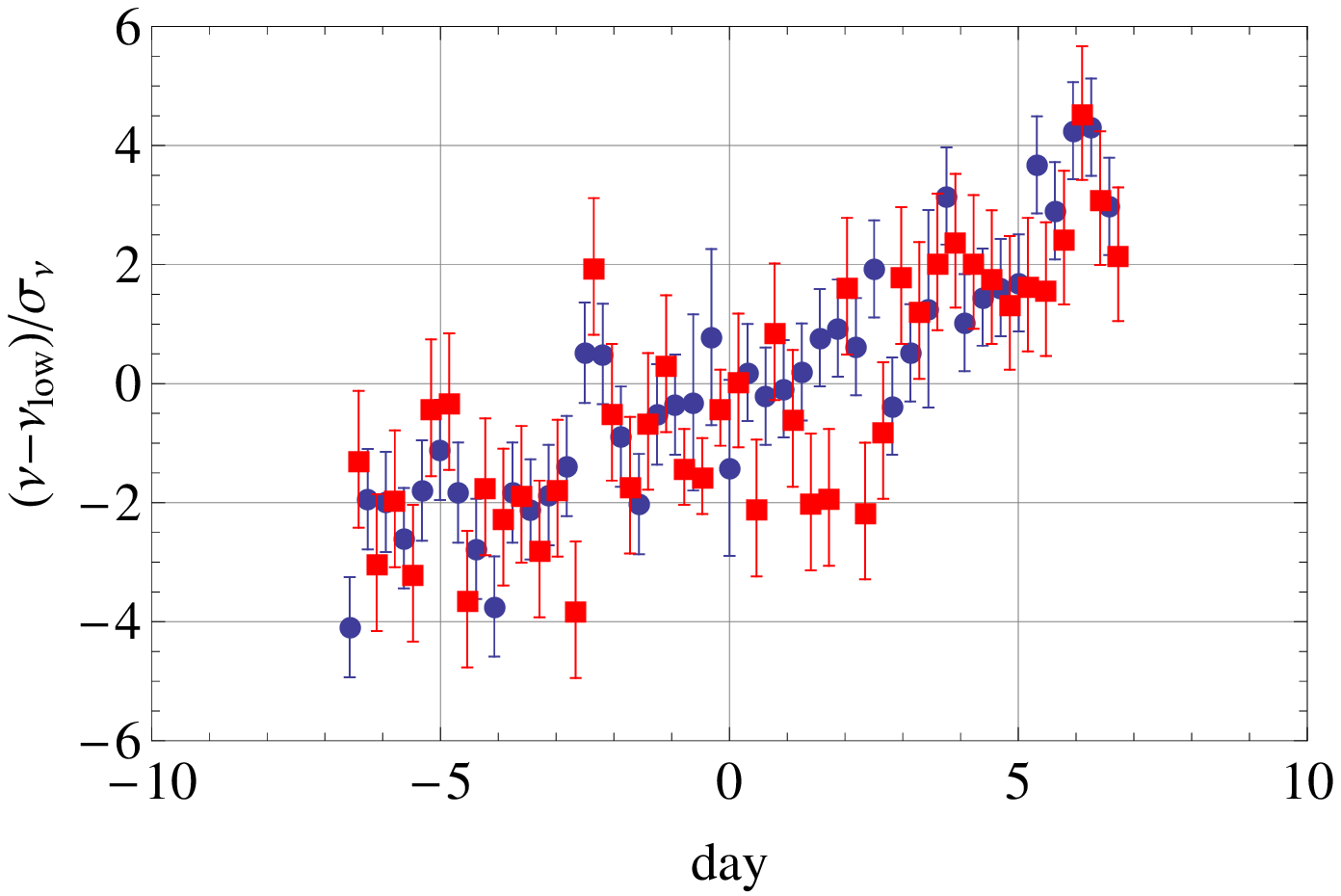}\hspace{-2mm}
\includegraphics[width=6.6cm]{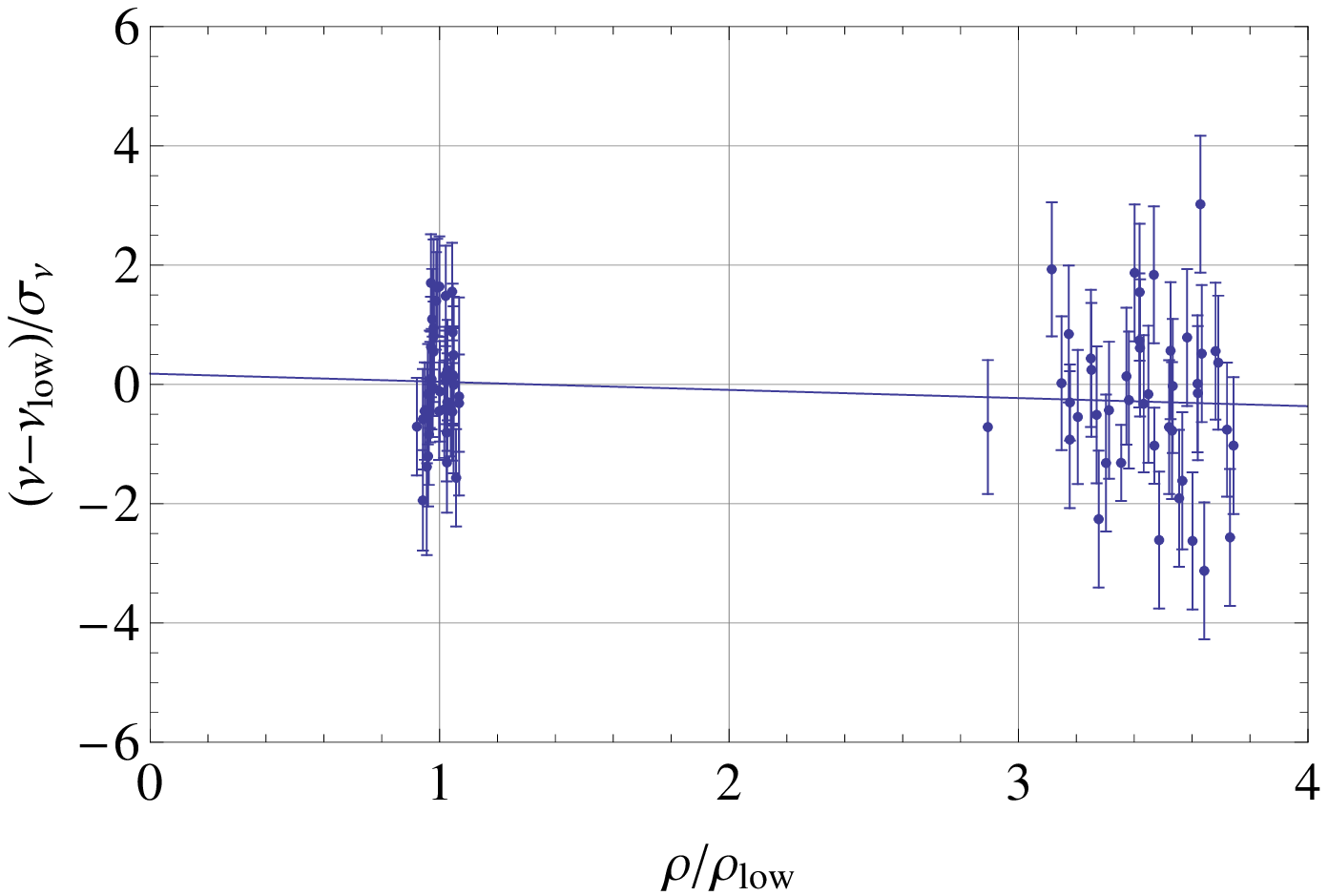}
\caption{Results of low (red squares) and high (blue bullets) density frequency measurements. The hydrogen-maser drift has been removed and the line is the intersection of the best-fit plane with $t=0$.} \label{data}
\end{figure}

Table 1 records measurement results and the relevant uncertainties, collected during a TAI unit evaluation run of the IT-CsF1 fountain at INRIM. The density values have been so scaled that the mean low-density is unitary; the frequency values have been given in units of the $\sigma_\nu = 3.9\times 10^{-15} \nu_{\rm Cs}$ uncertainty -- where $\nu_{\rm Cs} =$ 9 192 631 770 Hz -- and have been so shifted that the mean low-frequency is one. The same data are shown in Fig.\ \ref{data}.

\begin{figure}[b]\centering
\includegraphics[width=7cm]{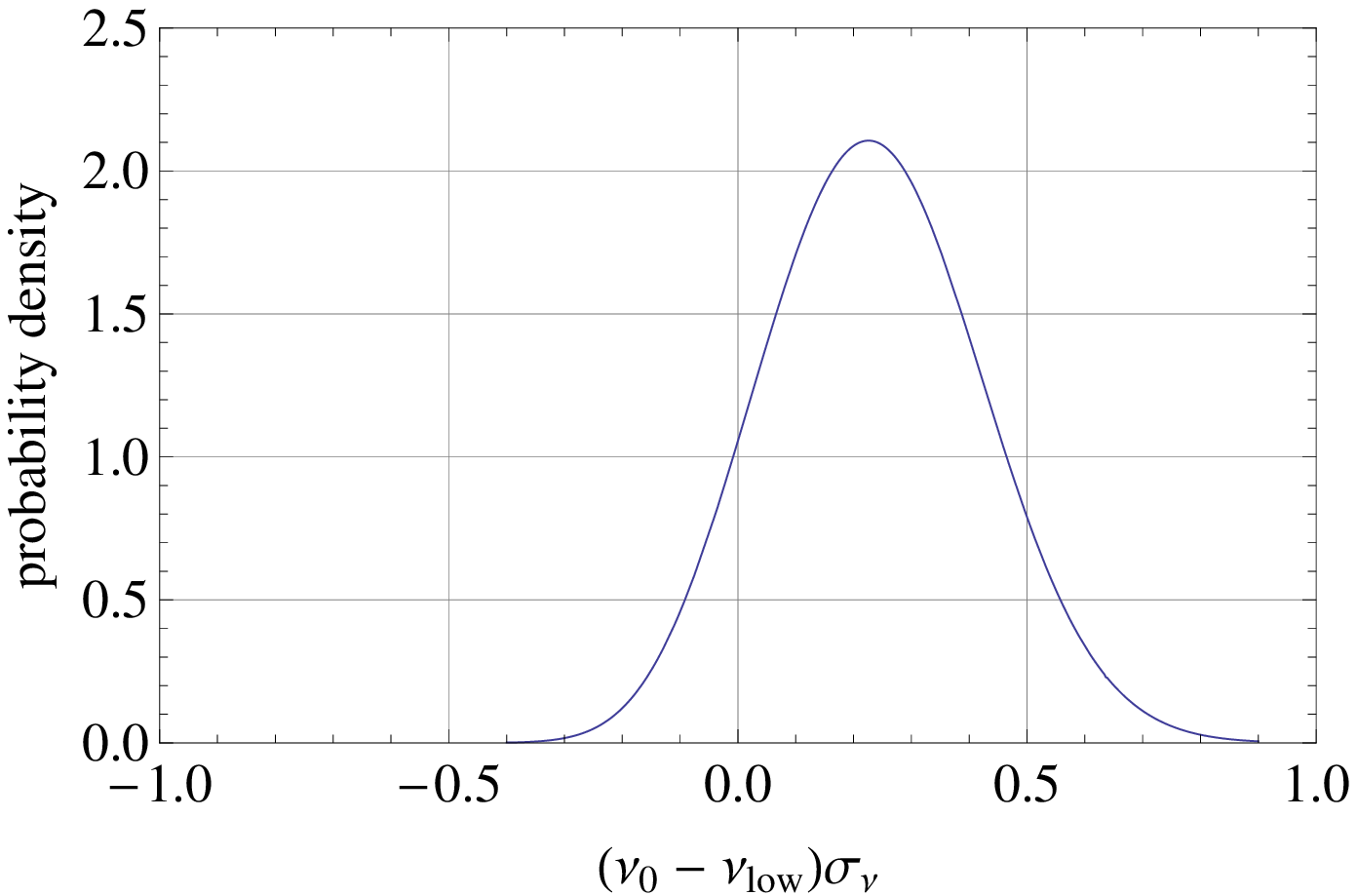}
\caption{Post-data probability density of the zero-density extrapolated frequency.} \label{f08}
\end{figure}

Some of the low- and high-density data pairs in Table 1 lie on a positive-slope line, in agreement with Gaussian dispersion of the frequency data. By the orthodox approach, the prior information is not taken into account and these positively-sloped data contribute to the best-fit line with the same weight as the others. It is not so when a Bayesian analysis is made. By applying the model described in \ref{sbs-rem} to the data in Table 1, the marginal post-data probability density of the zero-density frequency is given by (\ref{postdataeq}) and it is shown in Fig. \ref{f08}, where we used the $\tilde{c}=(0.41\pm 0.05)\sigma_\nu$/day pre-data estimate of the hydrogen maser drift.

After the relevant marginalization, the mean and standard deviation of the hydrogen maser drift, collisional coefficient, and zero-density frequency have been calculated and are shown in Table 2 \cite{math}. Since, in general, the probability densities of these quantities are not Gaussian, the mathematical definitions of the mean and variance, for instance,
\begin{equation}
 {\rm E}(b) = \int_{-\infty}^{+\infty} \xi P_b(\xi|\bi{y})\; \rmd \xi
\end{equation}
and
\begin{equation}
 {\rm var}(b) = \int_{-\infty}^{+\infty} \big[\xi-{\rm E}(b)\big]^2 P_b(\xi|\bi{y})\; \rmd \xi ,
\end{equation}
have been used. The same table shows also the classical estimates obtained by fitting the data with the $ax_i+b+ct_i$ plane.

When comparing the Bayesian and the orthodox frequency extrapolation, we observe that the Bayes theorem allows the orthodox uncertainty to be reduced by 28\%. This is in agreement with the best usage of the prior information, which usage reduces significantly the variability of the regression coefficient. In terms of original relative frequency units, the Bayesian analysis allows the extrapolation uncertainty to be reduced from $9.8\times10^{-16}\nu_{\rm Cs}$ to $7.0\times10^{-16}\nu_{\rm Cs}$. We also checked that the use of a more conservative pre-data estimate of the hydrogen maser drift, e.g., $(0.50\pm 0.15)\sigma_\nu$/day, does not significantly affect the post-data probability density.

\begin{table}[h]
\centering
\caption{Comparison of classical and Bayesian estimates of the hydrogen maser drift, collisional coefficient, and zero-density frequency. $\sigma_\nu=3.9\times 10^{-15} \nu_{\rm Cs}$ is the mean uncertainty of frequency data} \vspace{2mm}
\begin{tabular}{llll}
\hline
 &classical analysis &Bayesian analysis \\
\hline
 H maser drift           &$(0.43\pm 0.03)\sigma_\nu$/day              &$(0.44\pm 0.02)\sigma_\nu$/day \\
 collisional coefficient &$(-0.14\pm 0.10)\sigma_\nu/\rho_\low$       &$(-0.18\pm 0.08)\sigma_\nu/\rho_\low$ \\
 zero-density frequency  &$\nu_\low+(0.18\pm 0.25)\sigma_\nu$         &$\nu_\low+(0.24\pm 0.18)\sigma_\nu$ \\
\hline
\end{tabular}
\end{table}

A final consideration concerns the conceptual interpretation of classical and Bayesian results. When summarized by the orthodox best estimate and uncertainty, the available information about the zero-density frequency is synthesized by the realization of a random variable (the extrapolated frequency value) and by a measure of the width of distribution from which this realization has been sorted -- e.g., the variance. On the contrary, the Bayesian analysis synthesizes the information by a measure of the measurand-distribution location and by a coverage interval containing the value of the zero-density frequency with stated probability.

\section{Conclusions}
When extrapolating the frequency of a caesium-fountain clock to zero atom-density, the Bayes theorem makes it possible to take account of a negative correlation between the density and the frequency at the very beginning of data analysis, in a way ensuring logical consistency of the extrapolation.

In the specific example here considered, the theorem demonstrates itself capable of estimating the hydrogen maser drift and the collisional coefficient at the same time and of improving the orthodox extrapolation of the zero density frequency. With use of the sign information, the frequency extrapolation can be efficiently carried out also in the case of a single data pair and also when the measured frequencies are inconsistent, that is, when they violate the $y_2>y_1$ constraint.

We have considered the accuracy evaluation of the INRIM Cs fountain during standard operation; the uncertainty contribution due to atomic density shift is reduced by 28\%, from $9.8\times10^{-16}$ to $7.0\times10^{-16}$. In this application, the frequencies are averaged classically over 21000 s (low-frequency configuration) and 6000 s (high-frequency configuration); the Bayesian analysis is then applied to the resulting set of data. Since these averages do not account for a negative collisional coefficient, future work will be aimed at investigating the optimal split the 27000 s block into shorter ones to reduce the influence of the classical averaging.

\begin{figure}[b]\centering
\includegraphics[width=7cm]{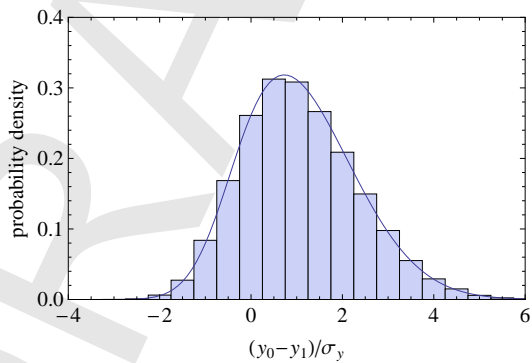}
\caption{Monte Carlo histogram of the zero-density frequency, given $\{x_1=1,y_1=0\}$ and $\{x_2=3,y_2=-1\}$. Solid line is the theoretical prediction (\ref{marginal})} \label{histo1}
\end{figure}

\appendix
\section{Inverse Monte Carlo simulation}
The probability density of the zero-density frequency (\ref{marginal}) was calculated also numerically by inverse Monte Carlo simulation. In direct Monte Carlo simulation a random list of measurement results, having a fixed measurand value, is generated by repetitions of a numerical experiment. On the contrary, in inverse simulation it is generated a random list of measurand values having a fixed measurement result.

A {\it brute force} approach exemplifies the simulation procedure. Let the measured values of the low- and high-density frequencies be fixed, for example, $y_1=0$ and $y_2=-1$ -- in the normalized units where $\sigma_y=1$. In the inverse simulation, $y_1$ and $y_2$ are sampled from $N(y_i|ax_i+b,1)$, where the slope and intercept are randomly chosen according (\ref{prior}). If the sampled frequencies are $y_1=0$ and $y_2=-1$, that is, if they happen to be identical to the wanted measurement results, the sorted zero frequency $b$ is appended to the Monte Carlo list.

This procedure is thoroughly inefficient. A short-cut is to observe that any arbitrary $\{y_1,y_2\}$ pair -- sorted from $N(y_i|ax_i+b,1)$, with arbitrarily chosen values of $a$ and $b$ -- can be mapped into the wanted $\{0,-1\}$ pair by
\begin{equation}
 y_i \rightarrow y_i - \frac{(1+y_2-y_1)(x_i-x_1)}{x_2-x_1} - y_1 .
\end{equation}
The same result should have been obtained if the data pair should have been sampled from $N(\nu_i|\nu_i,1)$, where
\begin{equation}\label{a2}
 \nu_i = \bigg( a-\frac{1+y_2-y_1}{x_2-x_1} \bigg) x_i + b + \frac{1+y_2-y_1}{x_2-x_1}x_1 - y_1 .
\end{equation}
Hence, given any $\{y_1,y_2\}$ sampled from $N(y_i|ax_i+b,1)$, the inverse Monte Carlo simulation sorts randomly the zero-frequency
\begin{equation}\label{a3}
 b \rightarrow b + \frac{y_2-y_1}{x_2-x_1}x_1 - y_1 ,
\end{equation}
which is obtained by setting $x_i=0$ in (\ref{a2}). Provided the relevant collisional coefficient in (\ref{a2}), $a-(1+y_2-y_1)/(x_2-x_1)$, is negative -- as requested, (\ref{a3}) is appended to the Monte Carlo list, otherwise it is rejected.

A Mathematica script illustrating the inverse Monte Carlo simulation is appended below; the simulation results are shown in Fig.\ \ref{histo1}: the Monte Carlo frequencies agree with the marginal probability density of the zero-density frequency predicted by (\ref{marginal}). We are unable to find out if and where the uniformity of the pre-data distribution has been used in the simulation; therefore, this result confirms that (\ref{prior}) indicates correctly the absence of any prior information, apart from the sign.

\begin{verbatim}
(* Inverse Monte Carlo Simulation *)
np = 100000;      (* # Monte Carlo runs *)
a  = -0.5; b = 0; (* model parameters, initially -0.5 and 0 *)
x1 =  1;  x2 = 3; (* low and high densities *)
(*  list of low density frequencies *)
y1 = a x1 + b + RandomReal[NormalDistribution[0, 1], {np}];
(* list of high density frequencies *)
y2 = a x2 + b + RandomReal[NormalDistribution[0, 1], {np}];
(* list of updated collisional coefficients *)
a  = a - (1 + y2 - y1)/(x2 - x1);
(* list of updated zero frequencies *)
b  = b + (1 + y2 - y1) x1/(x2 - x1) - y1;
ab = Transpose[{a, b}];
ab = Select[ab, #[[1]] < 0 &]; (* selects a < 0 values *)
b  = Transpose[ab][[2]];       (* Monte Carlo list *)
\end{verbatim}

\setcounter{section}{1}
\section{Sufficient statistics for the zero-density frequency and collisional coefficient}
The post-data probability density (\ref{post-multi}) holds because $\hat a$ and $\hat{y}$ -- the slope and intercept of the best-fit line through the data -- are sufficient statistics for the collisional coefficient and zero-density frequency. To demonstrate this, let the sampling distribution (\ref{many}) be written as
\begin{equation}\label{b1}
 P_\nu(\bi{y}|\bbeta) \propto \exp \bigg[ -\frac{1}{2}(\bi{y}-A\bbeta)^T C_y^{-1} (\bi{y}-A\bbeta) \bigg] ,
\end{equation}
where $\bi{y}$ is the vector of the measured frequencies, $\bbeta$ is the vector of the unknowns, $A$ is the design matrix, and $C_y$ the covariance matrix. The exponent of (\ref{b1}), $\chi^2(\bbeta)$, is a quadratic form in $\bbeta$; hence
\begin{equation}\label{b1}
 \chi^2(\bbeta) = {\rm const.} - \frac{1}{2}(\bbeta-\hat{\bbeta})^T C_{\bbeta}^{-1} (\bbeta-\hat{\bbeta}) ,
\end{equation}
where $\hat{\bbeta}=C_\beta A^T C_y^{-1} \bi{y}$ is the least-squares estimate of $\bbeta$ and $C_\beta = (A^T C_y^{-1} A)^{-1}$ its covariance matrix. Eventually, by leaving out the terms independent of $\bbeta$, which are unessential in (\ref{post-multi}),
\begin{equation}
 P_\nu(\bi{y}|\bbeta) \propto \exp \bigg[ -\frac{1}{2}(\bbeta-\hat{\bbeta})^T C_{\bbeta}^{-1} (\bbeta-\hat{\bbeta}) \bigg] .
\end{equation}

\end{document}